\documentclass[fleqn,usenatbib]{mnras}
\pdfoutput=1
\usepackage{graphicx}	
\usepackage{amsmath}	
\usepackage{amssymb}	
\usepackage{bm}
\newcommand{\kmax}{k_{\text{max}}}
\newcommand{\hMpcinv}{h\, \text{Mpc}^{-1}}
\newcommand{\hinvMpc}{h^{-1}\, \text{Mpc}}

\expandafter\def\expandafter\UrlBreaks\expandafter{\UrlBreaks
  \do\a\do\b\do\c\do\d\do\e\do\f\do\g\do\h\do\i\do\j%
  \do\k\do\l\do\m\do\n\do\o\do\p\do\q\do\r\do\s\do\t%
  \do\u\do\v\do\w\do\x\do\y\do\z\do\A\do\B\do\C\do\D%
  \do\E\do\F\do\G\do\H\do\I\do\J\do\K\do\L\do\M\do\N%
  \do\O\do\P\do\Q\do\R\do\S\do\T\do\U\do\V\do\W\do\X%
  \do\Y\do\Z}
  
\title[Cosmological Model Parameter Dependence of Matter Power Spectrum Covariance]{Cosmological Model Parameter Dependence of the Matter Power Spectrum Covariance from the DEUS-PUR {\it Cosmo} Simulations}

\author[Blot et al.]{
Linda Blot,$^{1}$\thanks{E-mail: lblot@mpa-garching.mpg.de}
Pier-Stefano Corasaniti,$^{2,3}$
Yann Rasera,$^{2}$
Shankar Agarwal$^{4}$
\\
$^{1}$Max-Planck-Institut f\"ur Astrophysik, Karl-Schwarzschild Str. 1, 85741 Garching, Germany \\
$^{2}$Laboratoire Univers et Th{\'e}ories, Observatoire de Paris, Universit{\'e} PSL, CNRS, Universit{\'e} de Paris,\\
5 place Jules Janssen 92190 Meudon, France \\
$^{3}$Sorbonne Universit\'e, CNRS, UMR 7095, Institut d'Astrophysique de Paris, 98 bis bd Arago, 75014 Paris, France\\
$^{4}$African Institute for Mathematical Sciences, 6 Melrose Road, Muizenberg, 7945, Cape Town, South Africa}

\begin{document}
\label{firstpage}
\pagerange{\pageref{firstpage}--\pageref{lastpage}}
\maketitle

\begin{abstract}
Future galaxy surveys will provide accurate measurements of the matter power spectrum across an unprecedented range of scales and redshifts. The analysis of these data will require one to accurately model the imprint of non-linearities of the matter density field. In particular, these induce a non-Gaussian contribution to the data covariance that needs to be properly taken into account to realise unbiased cosmological parameter inference analyses. Here, we study the cosmological dependence of the matter power spectrum covariance using a dedicated suite of N-body simulations, the Dark Energy Universe Simulation - Parallel Universe Runs (DEUS-PUR) {\it Cosmo}. These consist of 512 realizations for 10 different cosmologies where we vary the matter density $\Omega_m$, the amplitude of density fluctuations $\sigma_8$, the reduced Hubble parameter $h$ and a constant dark energy equation of state $w$ by approximately $10\%$. We use these data to evaluate the first and second derivatives of the power spectrum covariance with respect to a fiducial $\Lambda$CDM cosmology. We find that the variations can be as large as $150\%$ depending on the scale, redshift and model parameter considered. By performing a Fisher matrix analysis we explore the impact of different choices in modelling the cosmological dependence of the covariance. Our results suggest that fixing the covariance to a fiducial cosmology can significantly affect the recovered parameter errors and that modelling the cosmological dependence of the variance while keeping the correlation coefficient fixed can alleviate the impact of this effect.
\end{abstract}


\section{Introduction}
The upcoming generation of galaxy surveys will provide accurate measurements of the clustering of matter across an unprecedented range of scales and redshifts \citep[e.g.][]{2009arXiv0912.0201L,2011arXiv1110.3193L,2016arXiv161100036D,2019arXiv190205569A}. Precise estimates of the matter power spectrum from measurements of the spatial distribution of galaxies and the weak gravitational lensing shear will enable one to test models beyond the standard $\Lambda$CDM scenario and investigate the nature of dark energy. These datasets will be sensitive to the imprints of the non-linear regime of gravitational collapse of matter; as such they need to be accurately modelled if one aims to infer unbiased cosmological parameter constraints. In the past years, this has sparked a major theoretical and numerical effort to provide accurate predictions of galaxy clustering observables and the associated data covariances on quasi-linear and non-linear scales. On large scales the matter density field is Gaussian. Consequently, the matter power spectrum covariance has a diagonal structure and is simply proportional to the square of the power spectrum itself. Finite survey volume effects induce a non-Gaussian contribution also known as super-sample covariance \citep[see e.g.][]{2006MNRAS.371.1188H,2013PhRvD..87l3504T}, while the non-linearities of the matter density field that develop at small scales induce mode correlations that further contribute to the non-Gaussian structure of the covariance, making the matrix non-diagonal \mbox{\citep{1999MNRAS.308.1179M,1999ApJ...527....1S}}. Analytical approaches to estimate these effects have been developed in a vast literature \citep[see e.g.][]{2014MNRAS.445.3382M,2016PhRvD..93l3505B,2017JCAP...06..053B,2020arXiv200705504T}. Approximate numerical methods to simplify the evaluation of the covariance have also been introduced in several studies \citep[][]{2015MNRAS.454.4326P,2016MNRAS.456.2662F,2016MNRAS.460.1567P,2017MNRAS.466L..83J}. Nevertheless, N-body simulations remain the primary tool to investigate the imprint of non-linearities, whilst providing the necessary benchmark to test the validity of analytical model predictions \citep[see e.g.][]{2009ApJ...700..479T,2011MNRAS.414.2235K,2015MNRAS.446.1756B,2018MNRAS.478.4602K,2019arXiv190905273V}. 

Estimating the covariance with the level of accuracy that is required to correctly analyse future galaxy survey data demands sampling the matter power spectrum from a very large suite of N-body simulations. As an example, in \citet{2015MNRAS.446.1756B} we have estimated the covariance using $\sim 10^4$ independent N-body simulations and shown that non-linearities induce significant deviations from the Gaussian prediction on modes $k\gtrsim 0.25 \,\hMpcinv$ and at redshift $z<0.5$. Furthermore, by taking advantage of the large simulation suite, \citet{2016MNRAS.458.4462B} have shown that more than $>5000$ realizations are necessary to reduce the impact of sample covariance errors on the estimated cosmological parameter uncertainties to sub-percent level. 

In these studies the power spectrum covariance has been evaluated for a fixed fiducial cosmological model. However, the imprint of non-linearities on the matter power spectrum is cosmology dependent \citep[see e.g.][]{1999ApJ...521L...1M,2009JCAP...03..014C,2010MNRAS.401..775A}. Hence, it is natural to expect that such dependence extends to the non-Gaussian part of the matter power spectrum covariance. Neglecting the variation of the covariance with the cosmological model parameters can introduce spurious systematic errors in the parameter inference analysis. This has been investigated in the past in the context of weak lensing shear power spectrum measurements especially in relation to the super-sample covariance \citep{2009A&A...502..721E,2012ApJ...760...97L,2013A&A...551A..88C,2019OJAp....2E...3K,2019A&A...631A.160H} and several methodologies have been developed to extrapolate the cosmological dependence from a finite set of simulations \citep[see e.g.][]{2013JCAP...11..009M,2015JCAP...12..058W,2017MNRAS.465.4016R}. 

Here, we aim to specifically investigate the cosmological dependence of the matter power spectrum covariance due to small scale non-linearities. We will make use of a large ensemble of N-body simulations for several cosmological parameter configurations to compute the first- and second-order derivatives of the power spectrum covariance. Our intent is to determine the amplitude of such derivatives and perform a preliminary evaluation of their impact on cosmological parameter inference through a forecast analysis. Moreover, to facilitate further progress in the analytical modelling of the cosmological dependence of the power spectrum covariance, we have made publicly available the numerical simulation data used in the study presented here.

The paper is organized as follows. In Section ~\ref{Sec:Methodology} we describe the simulations set and the covariance estimator. In Section \ref{Sec:Results} we present our results on the cosmological dependence of the covariance and its impact on cosmological parameter inference analyses. In Section \ref{Sec:Conclusion} we present our conclusions.

\section{Methodology}
\label{Sec:Methodology}
\subsection{N-body Simulation Suite}
\subsubsection{Numerical Codes \& Simulation Pipeline}
Building upon the automated pipeline developed for the Dark Energy Universe Simulations - Parallel Universe Runs (DEUS-PUR) project \citep{2015MNRAS.446.1756B}, we have realized a large suite of N-body simulations for different sets of cosmological parameters. We refer to this simulation suite as DEUS-PUR {\it Cosmo}. 

In the following, we will briefly describe the simulation pipeline and we refer the interested readers to \citet{2015MNRAS.446.1756B} for a more detailed description. The cosmological parameters for the different runs are provided by the user through namelist files, while the sequential call to the various codes, from the computation of the input tables containing the cosmological functions to the post-processing of the simulations, is entirely automatised. For a given cosmological model the first step consists in computing the linear matter power spectrum using the code CAMB \citep{2000ApJ...538..473L} and solving the Friedmann equations using a dedicated code called NewDarkCosmos. The respective output tables are input to the code generating the initial conditions in the former case and the N-body solver in the latter case. Then Gaussian initial conditions are generated with an optimized version of the code MPGRAFIC \citep{2008ApJS..178..179P}. The simulations are run using an improved version of the Adaptive Mesh Refinement (AMR) N-body code RAMSES \citep{2002A&A...385..337T}, which uses a multigrid Poisson solver \citep{2011JCoPh.230.4756G}. Finally, halos are detected with the halo finder code pFoF \citep{2014A&A...564A..13R}, which is based on the friends-of-friends algorithm, while power spectra are computed using an optimized version of the code POWERGRID \citep{2008ApJS..178..179P}, which uses a Fast Fourier Transform (FFT) algorithm. The matter density field is estimated on a Cartesian grid (twice thinner than the coarse AMR grid) with a Cloud-in-Cell mass assignment scheme. To minimize the effect of aliasing, we exclude all modes beyond half the Nyquist frequency of the density grid. 

\begin{table}
    \centering
    \begin{tabular}{c|c|c|c|c|}
       ${\textrm{model}}$ & $\Omega_m$ & $\sigma_8$ & $h$ & $w$\\
       \hline
       1 & \textbf{1.0000} & 0.801 & 0.72 & -1.0 \\
       \emph{2} & \emph{0.2573} & \emph{0.801} & \emph{0.72} & \emph{-1.0} \\
       
       3 & 0.2573 & 0.801 & 0.72 & \textbf{-1.2} \\
       4 & 0.2573 & 0.801 & 0.72 & \textbf{-0.8} \\
       
       5 & 0.2573 & \textbf{0.700} & 0.72 & -1.0 \\
       6 & 0.2573 & \textbf{0.900} & 0.72 & -1.0 \\
       
       7 & \textbf{0.3100} & 0.801 & 0.72 & -1.0 \\
       8 & \textbf{0.2046} & 0.801 & 0.72 & -1.0 \\
       
       9 & 0.2573  & 0.801 & \textbf{0.67} & -1.0 \\
       10 & 0.2573 & 0.801 &\textbf{0.77} & -1.0 \\
    \end{tabular}
    \caption{Cosmological parameter values of the DEUS-PUR {\it Cosmo} simulated models with flat geometry. Model 1 is an Einstein-de Sitter model, model 2 is our fiducial $\Lambda$CDM model (in italic) with parameters set consistently to the WMAP-7 year data, while all other models differ for a variation of one of the parameter values (in bold).  Models 5-6 are characterized by a $\pm 13\%$ variation of $\sigma_8$ with respect to the fiducial value, models 7-8 by a $\pm 20\%$ variation of $\Omega_m$ and models 9-10 by a $\pm 7\%$ variation of $h$. Models 3 and 4 are flat wCDM models with a $\pm 20\%$ variation of the equation of state parameter with respect to the cosmological constant case ($w=-1$).}
    \label{tab:cosmo}
\end{table}

\begin{figure*}
\includegraphics{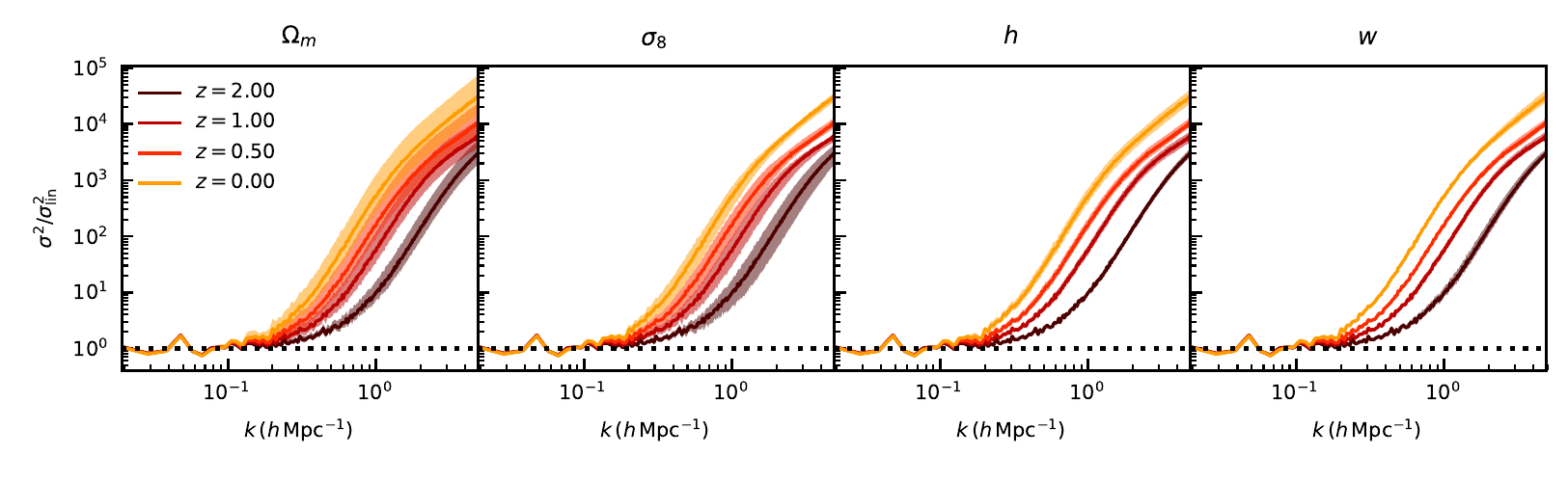}\\
\includegraphics{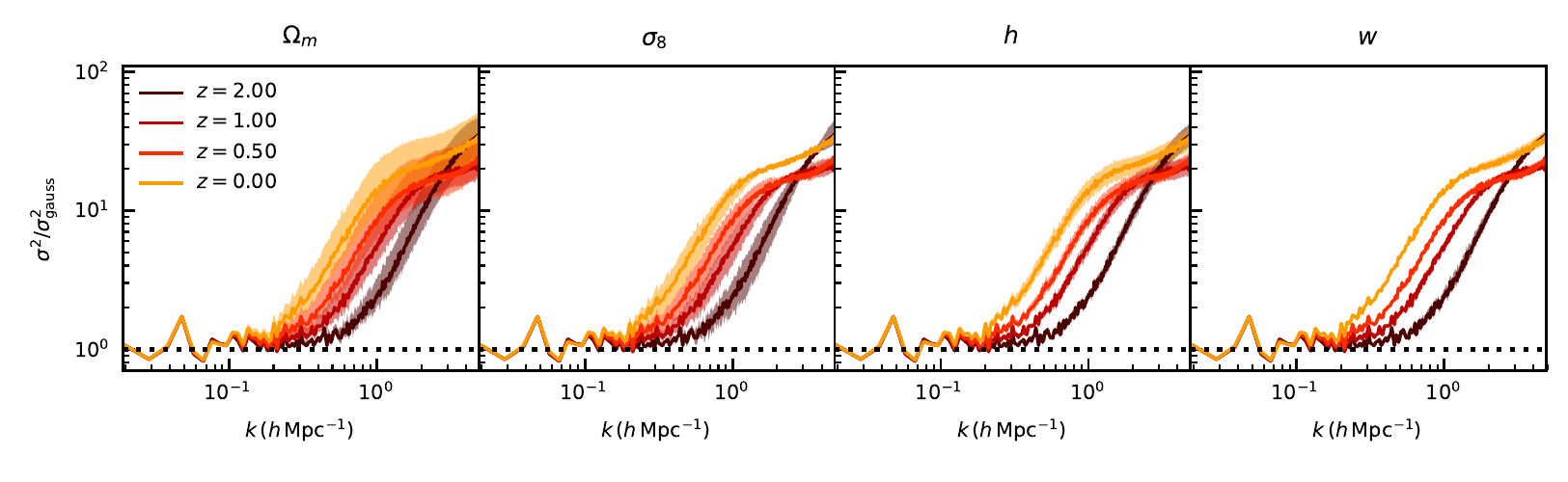}
\caption{Variance of the matter power spectrum as a function of wave number for different cosmologies normalized to the linear theory (top panels) and the Gaussian expectation (bottom panels). The continuous line indicates the fiducial cosmology case while the shaded area represents the variation when the parameter indicated in the title of each panel is varied. Colours from dark brown to yellow correspond to decreasing redshifts: $z=2$, $z=1$, $z=0.5$ and $z=0$. As we can see, non-linearities play an important role for $k>0.1-0.2\,\hMpcinv$ and are cosmology-dependent.}
\label{fig:variance}
\end{figure*}

\subsubsection{Cosmological Models \& Simulation Characteristics}

For each of the $10$ cosmological models listed in Table~\ref{tab:cosmo} we have run a set of $512$ cosmological N-body simulations that share the same initial phases across different cosmologies such as to reduce the numerical noise in the computation of the derivatives of the matter power spectrum and its covariance. Each simulation consists of a cubic box of length size $L_{\rm box}=328.125\, \hinvMpc$ with $512^3$ particles (corresponding to a particle mass $m_p=1.88\times10^{10}M_{\odot}/h$ for the fiducial cosmology). We have opted for such configuration, since it allows us to resolve with high accuracy the quasi-linear and non-linear scales contributing to the non-Gaussian part of the covariance, which is the object of our investigation.

 The cosmological parameters have been chosen such that we vary one parameter of interest at a time in a symmetric way with respect to the fiducial value. This allows us to obtain a very accurate estimate of the first and second derivative of any quantity in the vicinity of the reference cosmological model. Our fiducial cosmology corresponds to model 2 in Table~\ref{tab:cosmo}, which is a flat $\Lambda$CDM model calibrated on WMAP-7 year data \citep{2011ApJS..192...18K}. We have set the baryon density $\Omega_b h^2=0.02258$ and the scalar spectral index $n_s=0.963$ consistently with the values of the WMAP-7 cosmological analysis, while the amount of radiation and massless neutrinos is fixed to the standard values set by default in the CAMB code. Models 3 and 4 correspond to flat wCDM models with variations of the dark energy equation of state parameter $w$, model 5 and 6 correspond to variations of the amplitude of the matter density fluctuations on the $8 \,\hinvMpc$ scale $\sigma_8$, models 7 and 8 to variations of the cosmic matter density $\Omega_m$, while models 9 and 10 are associated with variations of the reduced Hubble constant $h$. We have also run as test case a set of simulations for model 1, which is an Einstein-de Sitter model (EdS) without dark energy. 

The cosmological parameter values listed in Table~\ref{tab:cosmo} cover a range that is much larger than the $1\sigma$ limits inferred from the Planck analysis of the Cosmic Microwave Background (CMB) anisotropy power spectra \citep{2018arXiv180706209P} for the concordance $\Lambda$CDM model. The reasons for such a choice are multiple. First of all, the Planck constraints significantly relax in the case of the wCDM model considered here such as to include within $1\sigma$ the different combinations of parameter values of our simulated models. Moreover, these are consistent with the bounds inferred from state-of-the-art cosmological analyses of large-scale structure data \citep[see e.g.][]{2018PhRvD..98d3526A} for which the study presented here is particularly relevant. Secondly, our choice also reflects the fact that a too small parameter interval would result in a noisy estimation of the covariance derivatives. Finally, the parameter variations we have considered span tensions on the values of $H_0$ and the parameter combination $S_8=\sigma_8(\Omega_m/0.3)^{0.5}$ which have arisen from the analysis of different cosmological probes. As an example, direct measurements of $H_0$ have resulted in a $5\sigma$ discrepancy with the CMB inferred value from the Planck analysis \citep{2018arXiv180706209P}. Similarly, the value of $S_8$ measured from weak lensing probes is consistently lower than the value inferred from the Planck data \citep{2020A&A...633A..69H,2018PhRvD..98d3526A,2019PASJ...71...43H}. The origin of such discrepancies has yet to be elucidated. Whether they are the result of unaccounted systematics or a real effect, the parameter variations considered here allow us to address their potential impact on power spectrum analyses through the cosmological dependence of the non-Gaussian covariance.

\subsection{Covariance Estimator and Parameter Derivatives}

We compute the matter power spectrum $P(k)$ of each realization in band powers of size $\Delta{k}=2\pi/L_{\rm box}$ in the range of modes $k_{\rm min}\le k\le k_{\rm max}$, where $k_{\rm min}=2\pi/L_{\rm box}$ and $k_{\rm max}=k_{\rm Ny}/2=\pi N_p^{1/3}/L_{\rm box}$ with $k_{\rm Ny}$ being the Nyquist frequency of the density grid of the Cloud-in-Cell algorithm that we use to estimate the spectra. The density grid is twice thinner than the coarse AMR grid of the simulation. In particular, we have $k_{\rm min}=\Delta{k}\approx 0.02\,h$ Mpc$^{-1}$ and $k_{\rm max}\approx 4.90\,h$ Mpc$^{-1}$.

We evaluate the covariance between two different modes using the unbiased sample covariance estimator:
\begin{equation}
C_{k_1,k_2}=\frac{1}{N_s-1}\sum_{i=1}^{N_s}\left[P_i(k_1) - \bar{P}(k_1)\right]\left[P_i(k_2) - \bar{P}(k_2)\right]
\end{equation}
where $N_s$ is the number of realizations, $P_i(k)$ is the matter power spectrum of the $i-$th realization and $\bar{P}(k)=\sum_{i=1}^{N_s}P_i(k)/N_s$ is the sample mean. 

We estimate the first and second derivatives of the power spectrum covariance with respect to the cosmological parameters for each pair of modes using the finite difference approximation:
\begin{align}
    \frac{\partial C_{k_1,k_2}}{\partial \theta}&\approx\frac{C_{k_1,k_2}(\hat{\theta}+\Delta\theta)-C_{k_1,k_2}(\hat{\theta}-\Delta\theta)}{2\Delta\theta},\\
    \frac{\partial^2 C_{k_1,k_2}}{\partial \theta^2}&\approx\frac{C_{k_1,k_2}(\hat{\theta}-\Delta\theta)-2C_{k_1,k_2}(\hat{\theta})+C_{k_1,k_2}(\hat{\theta}+\Delta\theta)}{\Delta\theta^2},
\end{align}
where $\hat{\theta}$ is the fiducial cosmological parameter value and $\Delta\theta$ the finite variation of its value.

\begin{figure*}
\centering
\begin{tabular}{c}
\includegraphics[scale=0.95]{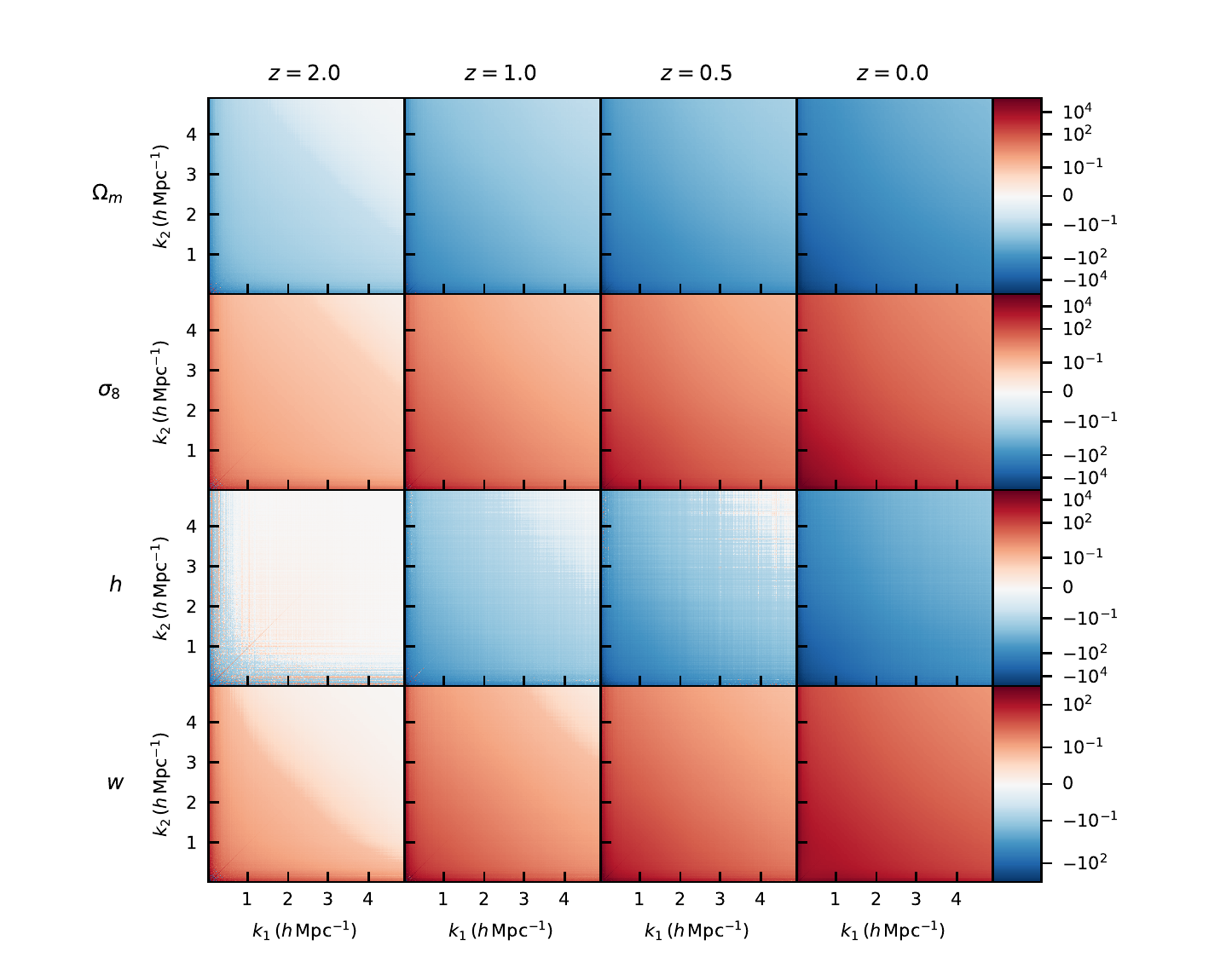}\\
\includegraphics[scale=0.95]{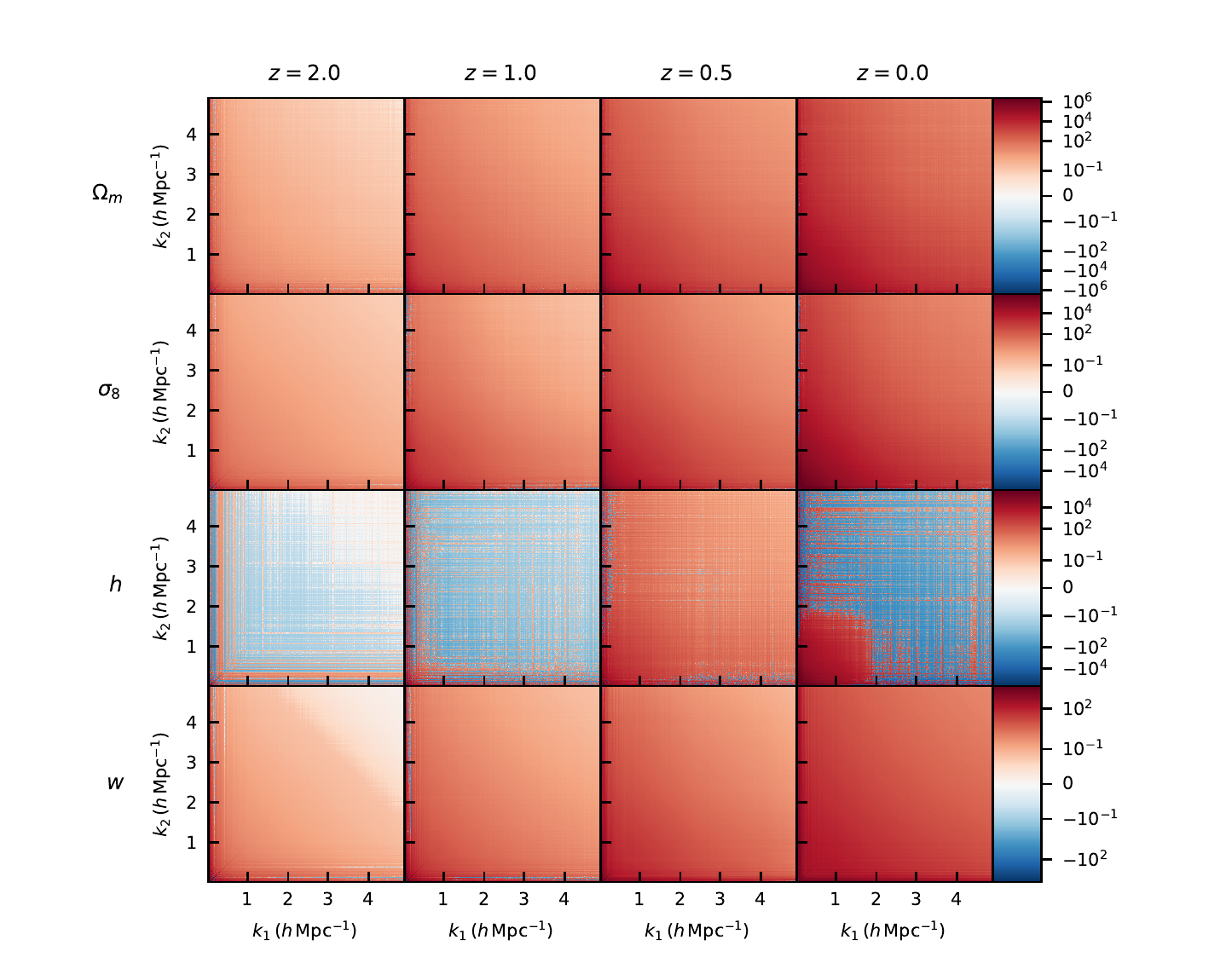}
\end{tabular}
\caption{First (top) and second (bottom) derivative of the covariance with respect to $\Omega_m$ (first row), $\sigma_8$ (second row), $h$ (third row) and $w$ (fourth row). The columns from left to right corresponds to redshift $z=2,1,0.5$ and $0$ respectively.}
\label{fig:cov_der}
\end{figure*}

\begin{figure*}
\centering
\begin{tabular}{c}
\includegraphics[scale=0.95]{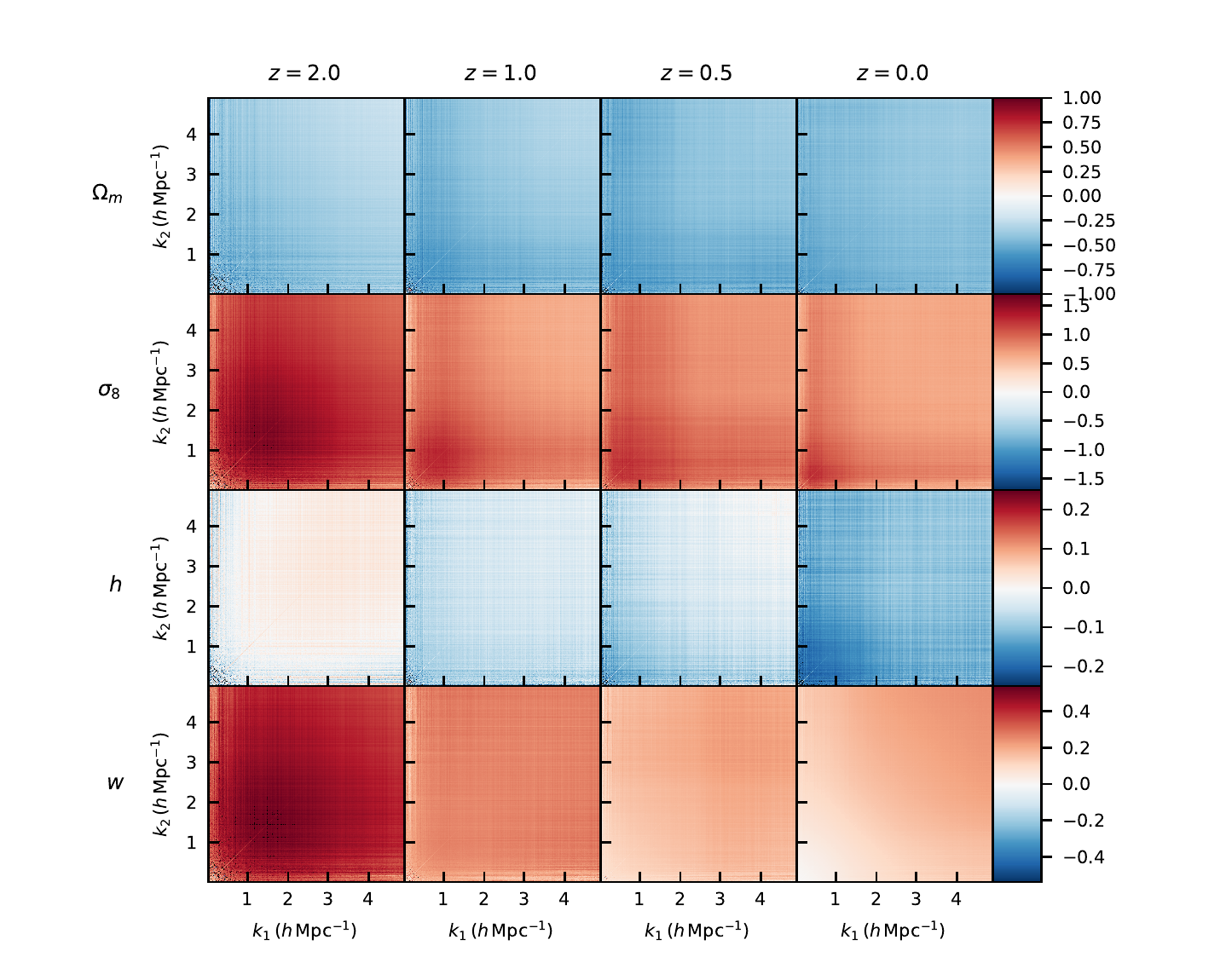}\\
\includegraphics[scale=0.95]{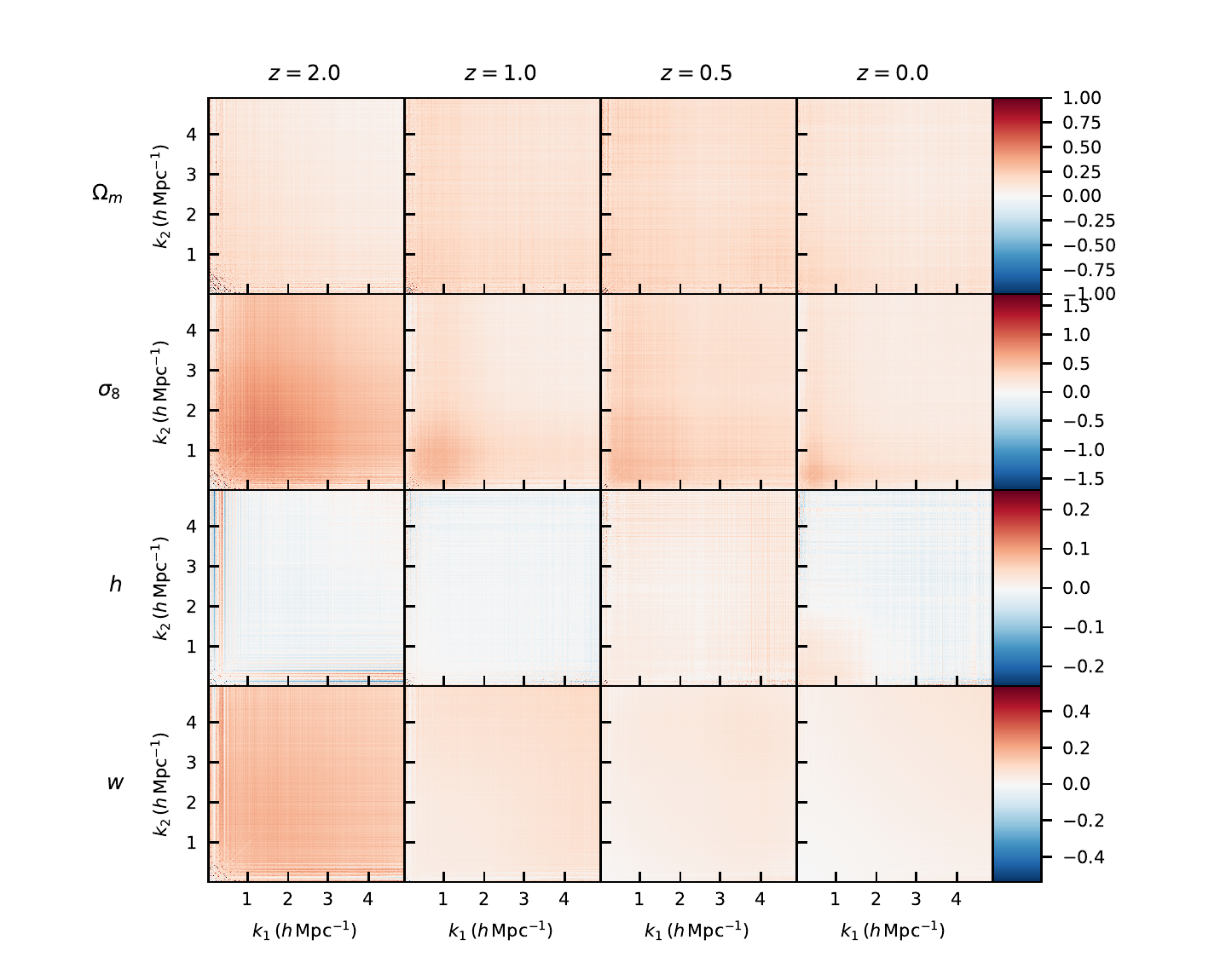}
\end{tabular}
\caption{First-order (top) and second-order (bottom) term of the Taylor expansion of the covariance (Eq.~\ref{eq:taylor}) normalized to the fiducial model covariance for the cosmological model parameters and redshifts as shown in Fig.~\ref{fig:cov_der}. Here we use the same color coding for both panels to highlight the relative importance of the expansion terms. Black pixels correspond to masked elements exceeding the boundary values due to sample variance noise. This shows the large variation of the covariance induced by small variations of the parameters: $\Omega_m (\pm20\%)$, $\sigma_8(\pm13\%)$, $h(\pm7\%)$, $w(\pm20\%)$.}
\label{fig:taylor}
\end{figure*}

\begin{figure*}
\centering
\includegraphics{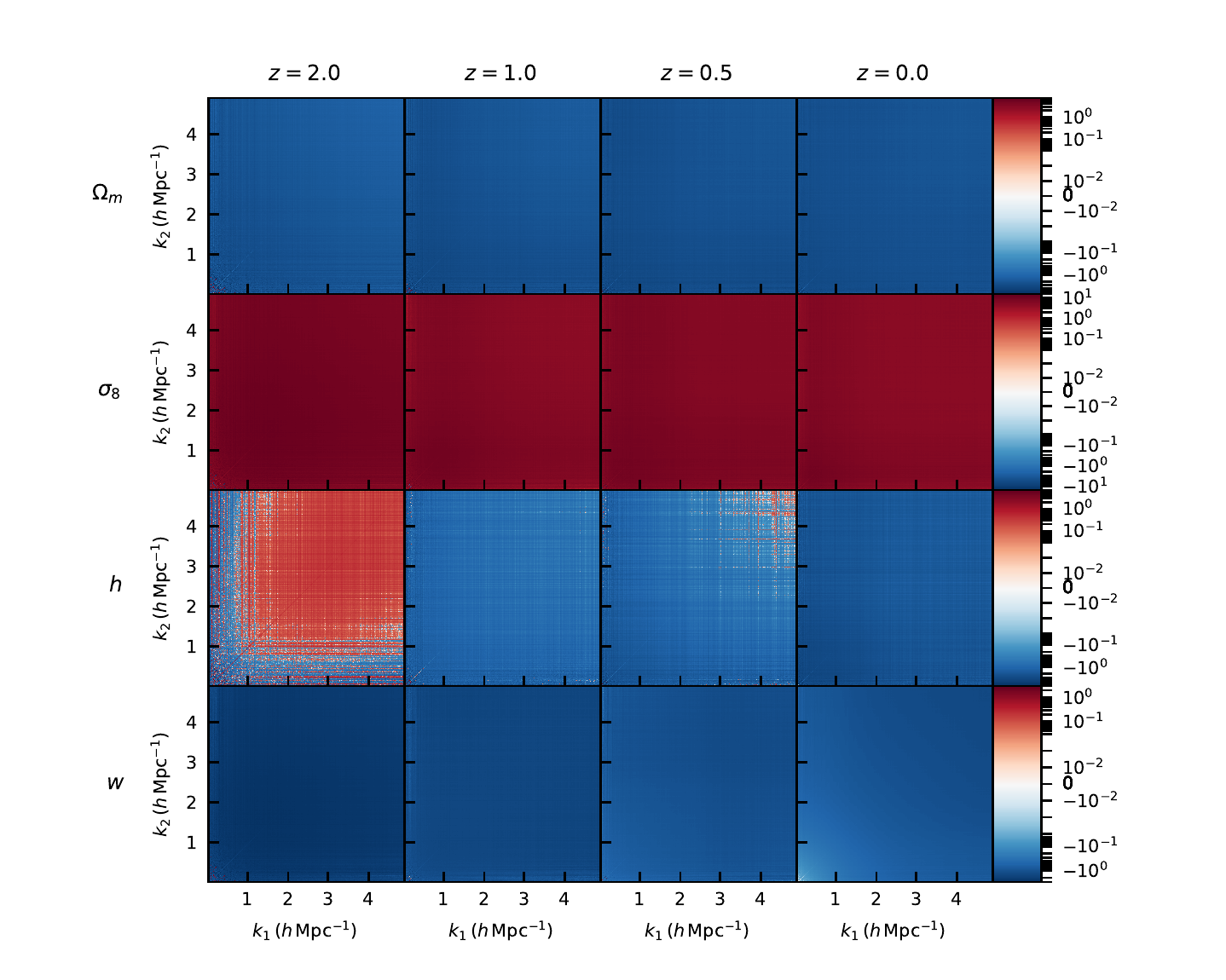}
\caption{Fractional variation of the covariance with respect to $\Omega_m$ (first row), $\sigma_8$ (second row), $h$ (third row) and $w$ (fourth row). The columns from left to right corresponds to redshift $z=2,1,0.5$ and $0$ respectively.}
\label{fig:log_der}
\end{figure*}

\section{Results}
\label{Sec:Results}

\subsection{Variance of the Matter Power Spectrum}
We evaluate the variance of the matter power spectrum (i.e. the diagonal part of the covariance matrix). This is shown in Figure~\ref{fig:variance} for the different cosmological models (panel left to right) and redshifts (lines from yellow to dark brown). The top panels show the variance normalized to the linear prediction, 
\begin{equation}
\sigma_{\rm lin}^2(k)=\frac{2\,P_{\rm lin}^2(k)}{N_{\rm modes}},\nonumber
\end{equation}
where $N_{\rm modes}\approx k^2\Delta{k}V/(2\pi^2)$ is the number of modes in the bin of width $\Delta k$ over the spatial volume $V$ and $P_{\rm lin}(k)$ is the linear matter power spectrum; the lower panels show the variance normalized to the Gaussian prediction 
\begin{equation}
\sigma_{\rm gauss}^2(k)=\frac{2\,\bar{P}^2(k)}{N_{\rm modes}},\nonumber
\end{equation}
where $\bar{P}$ is the average non-linear matter power spectrum from the 512 independent N-body realizations. 

We may notice that at small wavenumbers ($k<0.1 \,\hMpcinv$) the estimated variance is consistent with the linear Gaussian prediction, which validates the results of our simulations in this regime. On the other hand at larger wavenumbers, we observe a strong departure that increases as function of the wavenumber and for decreasing redshifts. The amplitude of this deviation reaches up to a factor of $10^4$ greater than the linear prediction (top panels) at $k\approx 3\,\hMpcinv$ at $z=0$, and a factor $25$ with respect to the Gaussian case (bottom panels). In the first case this is partly due to the fact that the linear power spectrum significantly underestimates $P(k)$ at these scales and redshifts, while in the latter case this is due to the well-known non-Gaussian contribution from the non-linear regime of matter clustering \citep[see e.g.][and references therein]{2015MNRAS.446.1756B}. In all cases, we can see that the amplitude and slope of the departures from the linear and Gaussian expectations depend on the cosmological parameters in a non-trivial way. This motivates a detailed study of the cosmological dependence of the covariance which we discuss next.

\subsection{Matter Power Spectrum Covariance Derivatives}
We evaluate the first and second derivatives of the power spectrum covariance with respect to the cosmological parameters which we plot in the top and bottom panels of Fig.~\ref{fig:cov_der} respectively. Panels from left to right show the redshift evolution at $z=2,1,0.5$ and $0$. The intensity mapping is set such that positive (negative) derivatives are shown in red (blue), while vanishing matrix elements are shown in white. Panels from top to bottom correspond to derivatives with respect to $\Omega_m$, $\sigma_8$, $h$ and $w$ respectively. 

First, we may notice that the sign of the derivatives follows from that of the matter power spectrum. As an example, the first derivative of the covariance with respect to $\Omega_m$ is negative. This is because at a fixed value of $\sigma_8$, a positive variation of $\Omega_m$ decreases the overall amplitude of the matter power spectrum. Hence the covariance decreases, which results in a negative derivative. Conversely, a positive variation of $\sigma_8$ at a fixed $\Omega_m$ value increases the overall amplitude of matter power spectrum, thus resulting in a positive derivative. We can also see that the first-order derivative of the covariance increases in absolute value from high to low redshift. Moreover, at a given redshift the largest elements are those corresponding to pairs of modes consisting of a large mode coupled to a small one, which is indicative of the onset of non-linearities that grow at lower redshifts while propagating to larger scales. This leads to a characteristic off-diagonal structure of the first-order derivative of the covariance that similar to that of the covariance itself \citep[see e.g. Fig. 3 in][]{2015MNRAS.446.1756B}. It is worth noticing that the first-order derivative of the covariance is larger for $\Omega_m$ and $\sigma_8$ and smallest for $w$, which follows from the dependence of the matter power spectrum on these parameters. We observe a similar trend in the case of the second-order derivatives, shown in Fig.~\ref{fig:cov_der}, which all have positive values except for the case of $h$. Here, it is worth noticing that the derivatives with respect to $h$ have the smallest amplitude compared to the other parameters. Because of this, they are more sensitive to sample noise. This is particularly the case of the second derivative as noticeable from the bottom panel of Fig.~\ref{fig:cov_der}.

We can use these derivatives to infer an understanding of the dependence of the matter power spectrum covariance on the cosmological parameters. In particular, we can consider a Taylor expansion of the covariance up to second-order around the fiducial cosmology:
\begin{equation}\label{eq:taylor}
C_{k_1,k_2}(\theta)\approx C_{k_1,k_2}(\hat{\theta})+\frac{\partial C_{k_1,k_2}}{\partial \theta}\biggr|_{\hat{\theta}}(\theta-\hat{\theta})+\frac{1}{2}\frac{\partial ^2 C_{k_1,k_2}}{\partial \theta^2}\biggr|_{\hat{\theta}}(\theta-\hat{\theta})^2,
\end{equation}
the validity of this approximation depends on the expansion coefficients, i.e. the covariance derivatives normalised to the fiducial covariance, to be $\ll \mathcal{O}(1)$. We show these ratios in Fig.~\ref{fig:taylor} for the first-order (top panel) and second-order (bottom panel) terms respectively. In the first-order case we can see that the largest matrix elements are less than unity for $\Omega_m$, $h$ and $w$, though still large enough to cause a slow convergence of the Taylor expansion along these parameter directions. Instead, in the case of $\sigma_8$ the largest matrix elements exceed unity even on quasi-linear scales corresponding to modes $k\ll 1\,\hMpcinv$. This suggests that the dependence of the covariance on $\sigma_8$ may be highly non-linear. Using the same colour coding, we can see that the second-order terms are much smaller than the first-order ones. Moreover, most of the second-order contributions are smaller than unity, meaning that an important part of the information about the cosmological dependence of the covariance is encoded in the first two derivatives. In any case, it is striking that a variation of order $\simeq 10\%$ of the cosmological parameters induces a change of the covariance matrices between $10\%$ and $150\%$ depending on cosmology, redshift and scale.

In Fig.~\ref{fig:log_der} we also plot the fractional variation of the covariance, $\partial \log C_{k_1,k_2}/\partial \log \theta$. This allows to estimate the expected variation of the covariance when the parameters are varied by different amounts than the ones used in this work.

\subsection{Cosmological Parameter Inference Forecast}
In order to assess the impact of a cosmology dependent covariance on the cosmological parameter inference we perform a simple Fisher matrix analysis. To take into account the loss of information due to estimating the covariance from a finite number of simulations we employ Eq. (17) of \citet{2017MNRAS.464.4658S} and multiply all Fisher matrices by the factor:
\begin{equation}
\frac{N_s (N_s-N_b)}{(N_s-1)(N_s+2)},
\end{equation}
where $N_b$ is the number of bins in the data vector.

In principle, given the non-Gaussian structure induced by the cosmological parameter dependence of the covariance, a more rigorous approach would be to perform a Markov Chain Monte Carlo analysis of a set of synthetic matter power spectrum data for our fiducial cosmology and let the covariance vary along the random sampling of the cosmological parameter space. However, given the limited number of parameter configurations for which we have evaluated the covariance and the potentially non-linear nature of its parameter dependence, we are unable to perform such an analysis. 

\begin{figure}
\includegraphics{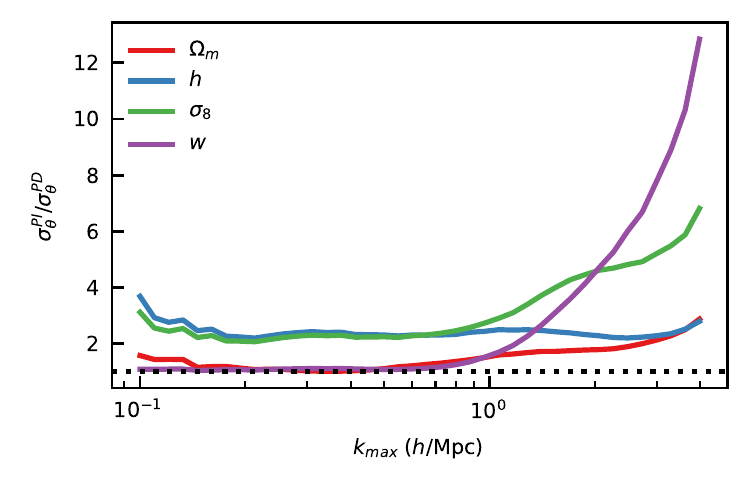}
\caption{Ratio of the parameter errors obtained with a parameter independent and a parameter dependent covariance matrix using Eqs.~\ref{eq:fisher_pi} and \ref{eq:fisher_pd} respectively.}
\label{fig:full_fisher}
\end{figure}

In the following, we first perform a parameter error forecast assuming the Fisher matrix defined as \citep{1997ApJ...480...22T}:
\begin{multline}\label{eq:fisher_pd}
    F^{PD}_{\alpha\beta}=\sum_{l}\left[\sum_{ij} \frac{\partial P(k_i,z_l)}{\partial \theta_{\alpha}} \biggr|_{\theta_{\alpha}=\hat{\theta}_{\alpha}}  \frac{\partial P(k_j,z_l)}{\partial \theta_{\beta}}\biggr|_{\theta_{\beta}=\hat{\theta}_{\beta}} C_{k_i,k_j}^{-1}(z_l) \right.\\
    \left. + \frac{1}{2} \text{tr}\left(  C^{-1}(z_l) \frac{\partial C(z_l)}{\partial \theta_{\alpha}} C^{-1}(z_l) \frac{\partial C(z_l)}{\partial \theta_{\beta}}  \right)\right],
\end{multline}
where $P(k,z)$ is the non-linear matter power spectrum estimated from the N-body simulation of our fiducial cosmology. Notice that the second term in the above equation accounts for the parameter dependence of the covariance. We evaluate the expected parameter errors as function of $\kmax$. Then, we compare the results to the case where the covariance is fixed to the fiducial cosmology resulting in the Fisher matrix to be given by:
\begin{equation}\label{eq:fisher_pi}
    F^{PI}_{\alpha\beta}=\sum_{ijl}\frac{\partial P(k_i,z_l)}{\partial \theta_{\alpha}} \biggr|_{\theta_{\alpha}=\hat{\theta}_{\alpha}}  \frac{\partial P(k_j,z_l)}{\partial \theta_{\beta}}\biggr|_{\theta_{\beta}=\hat{\theta}_{\beta}} C_{k_i,k_j}^{-1}(z_l).
\end{equation}
Here PD and PI stand for parameter dependent and parameter independent covariance respectively. We plot the ratio of the estimated cosmological parameter errors obtained in the two cases in Figure~\ref{fig:full_fisher}. We can see that neglecting the parameter dependence of the covariance can lead to over-estimating the parameter errors by a factor of a few for $\kmax< 1$ and up to a factor of $\sim10$ at higher $\kmax$.

{It is worth noticing that} even though varying the covariance during the sampling would be the correct Bayesian way to infer model parameter constraints from datasets with parameter dependent covariances \citep[see e.g.][]{2016MNRAS.463.1651M}, this is not how current matter power spectrum analyses are performed, since the data covariance is fixed at a fiducial cosmology. If the fiducial model is far from the true cosmology, then the constraints can be biased and the errors misestimated. Therefore, to evaluate the extent to which this affects the cosmological parameter inference we compute the Fisher matrix:
\begin{equation}\label{eq:fisher}
    F_{\alpha\beta}=\sum_{ijl}\frac{\partial P(k_i,z_l)}{\partial \theta_{\alpha}} \biggr|_{\theta_{\alpha}=\hat{\theta}_{\alpha}}  \frac{\partial P(k_j,z_l)}{\partial \theta_{\beta}}\biggr|_{\theta_{\beta}=\hat{\theta}_{\beta}} C_{k_i,k_j}^{-1}(z_l)\biggr|_{\boldsymbol{\theta}=\boldsymbol{\theta_*}},
\end{equation}
where we have considered the parameter vector $\boldsymbol{\theta}=\{\Omega_m, \sigma_8, w, h\}$. We compute the derivative of the matter power spectrum for the fiducial cosmology specified by the vector of values $\boldsymbol{\hat{\theta}}=\{0.2573,0.801,-1,0.72\}$ with the finite difference approximation using the spectra from the simulations, while we use the covariance matrix of the simulated cosmologies specified by the vector of values $\boldsymbol{\theta_*}$ from Table~\ref{tab:cosmo}. We evaluate Eq.~(\ref{eq:fisher}) assuming $15$ uncorrelated redshift bins in the range $0.15<z<1.75$ corresponding to the redshift of the snapshots of our simulation suite. 

\begin{figure*}
    \includegraphics{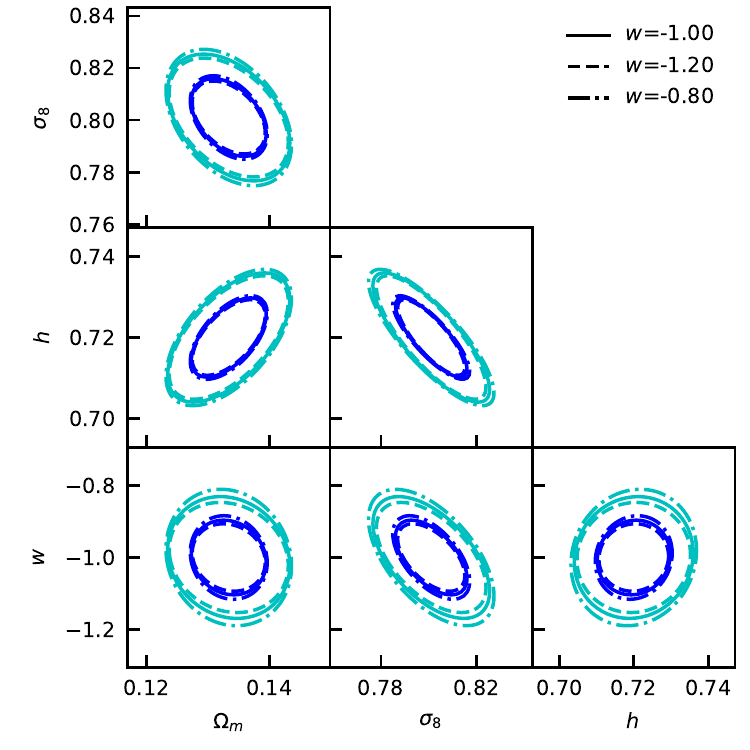}
    \includegraphics{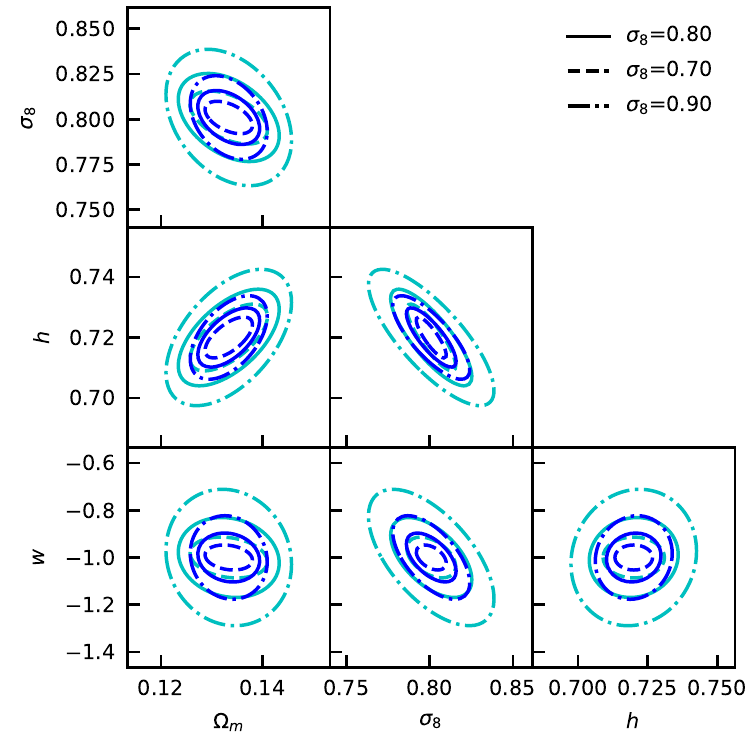}
    \caption{Examples of $1\sigma$ (dark blue) and $2\sigma$ (light blue) parameter contours computed from the Fisher matrix analysis assuming the covariance of the fiducial cosmology (solid lines), and that associated to a model with a non-fiducial parameter value (dashed and dot-dashed lines) of $w$ (left panel) and $\sigma_8$ (right panel).}
    \label{fig:contours}
\end{figure*}

In Figure~\ref{fig:contours} we show two examples of the $1$ and $2\sigma$ contours inferred from the evaluation of the Fisher matrix assuming $\kmax=1\,\hMpcinv$ and obtained using the covariance evaluated for different non-fiducial values of $\sigma_8$ (left panel) and $w$ (right panel). The contours inferred with the covariance evaluated at the fiducial cosmology are shown as solid line. We can see that with respect to this case, setting the covariance to a different cosmological model leads to a modification of the area within the confidence regions as well as the angle of the degeneracies between different pairs of parameters.

\begin{figure}
    \includegraphics{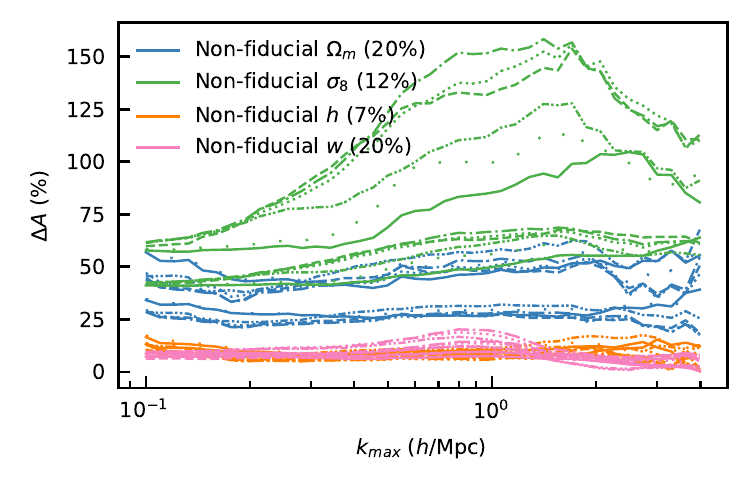}
    \includegraphics{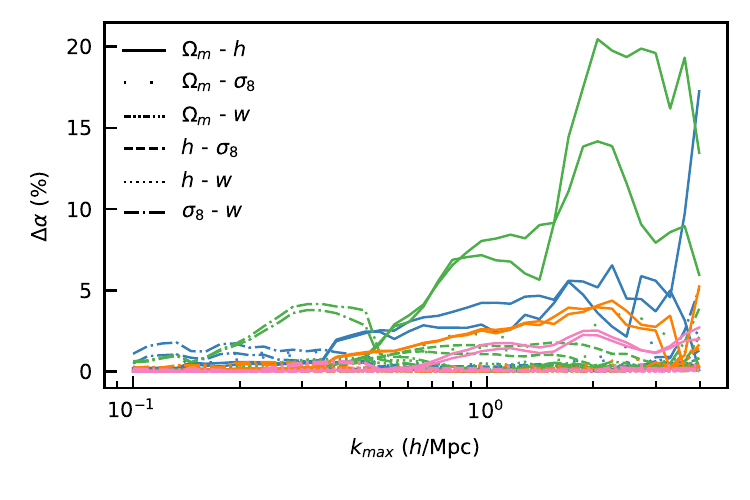}
    \caption{Variation of the area (top panel) and angle (bottom panel) of the 1$\sigma$ contours from the Fisher matrix analysis as a function of $\kmax$ when using the covariance with one of the parameters (indicated in the legend of the top panel) that is different from the fiducial one. The various line styles corresponds to the parameter pairs indicated in the legend of the bottom panel. For each pair there are two lines, corresponding to the positive and negative variation of the non-fiducial parameter.}
    \label{fig:area-angle}
\end{figure}

\begin{figure*}
    \includegraphics{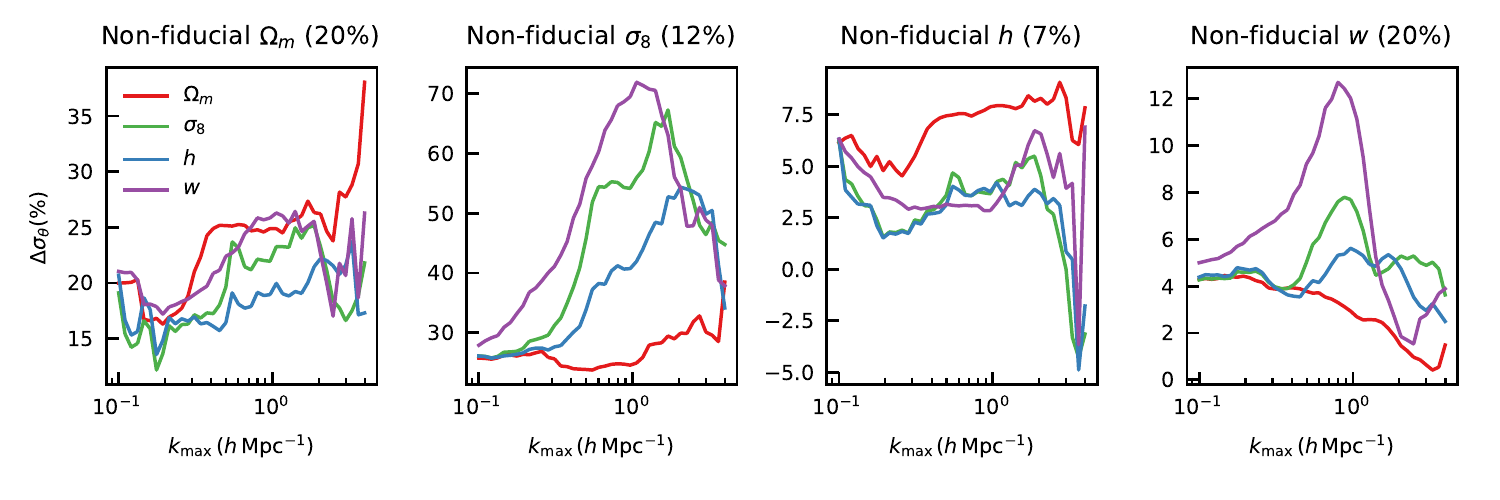}
    \caption{Variation of the 1$\sigma$ error from the Fisher matrix analysis as a function of $\kmax$ when using the covariance with one of the parameters (indicated in the title of each panel) that is different from the fiducial one.}
    \label{fig:param-error}
\end{figure*}

In Figure~\ref{fig:area-angle} we show the variation of the area (left panel) and the angle (right panel) of the 1$\sigma$ contours for different combination of parameters as a function of $\kmax$ when using a covariance computed for the non-fiducial cosmological models from Table~\ref{tab:cosmo} with different values of $\Omega_m$ (blue lines), $h$ (orange lines), $\sigma_8$ (green lines) and $w$ (pink lines). We can see that the largest deviation of the contour area occurs in the case of the covariance being computed for non-fiducial values of $\sigma_8$ and $\Omega_m$, while differences are smaller for $h$ and $w$. Quite importantly, deviations are of the order of $50\%$ on quasi-linear scales corresponding to $\kmax\sim 0.1-0.2 \,\hMpcinv$, which are already probed by current galaxy surveys \citep[see e.g.][]{2017MNRAS.466.2242B}. The effect on the angle of the parameter degeneracies is smaller with maximal deviations not exceeding the $5\%$ level up to $\kmax\sim 2 \,\hMpcinv$. For higher $\kmax$ the largest impact is associated with $\sigma_8$, while the effect remains smaller for the other parameters. 

In Figure~\ref{fig:param-error} we show the variation of the 1$\sigma$ parameter errors as a function of $\kmax$ when using a covariance in a cosmology with one of the parameters set to a non-fiducial value (indicated in the title of each panel). As already noted above, the most dramatic variations occur for non-fiducial values of $\sigma_8$ and $\Omega_m$. Quite remarkably, all the parameter errors considered here are already affected at low $\kmax$ values. 

\begin{figure}
    \includegraphics{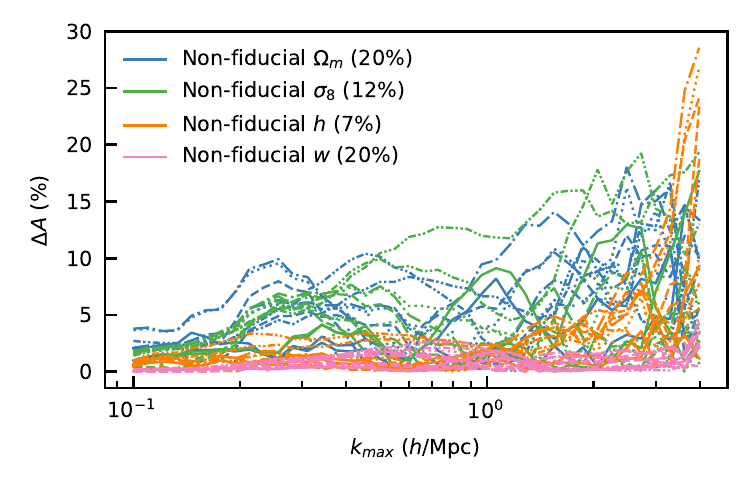}
    \includegraphics{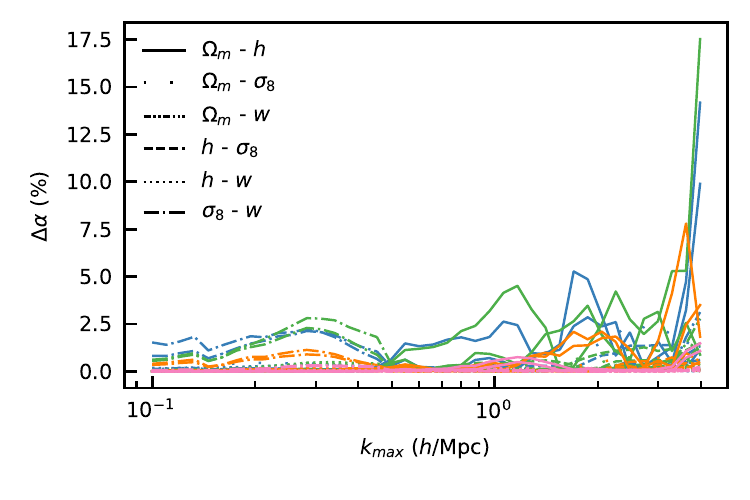}
    \caption{Variation of the area (top panel) and angle (bottom panel) of the 1$\sigma$ contours from the Fisher matrix analysis as a function of $\kmax$ assuming that the diagonal part of the covariance is that of the fiducial cosmology, while the off-diagonal elements are set to a cosmology with a non-fiducial value for the cosmological parameter indicated in the legend of the top panel. The various line styles corresponds to the parameter pairs indicated in the legend of the bottom panel. For each pair there are two lines, corresponding to the positive and negative variation of the non-fiducial parameter.}
    \label{fig:area-angle-diag}
\end{figure}

\begin{figure*}
    \includegraphics{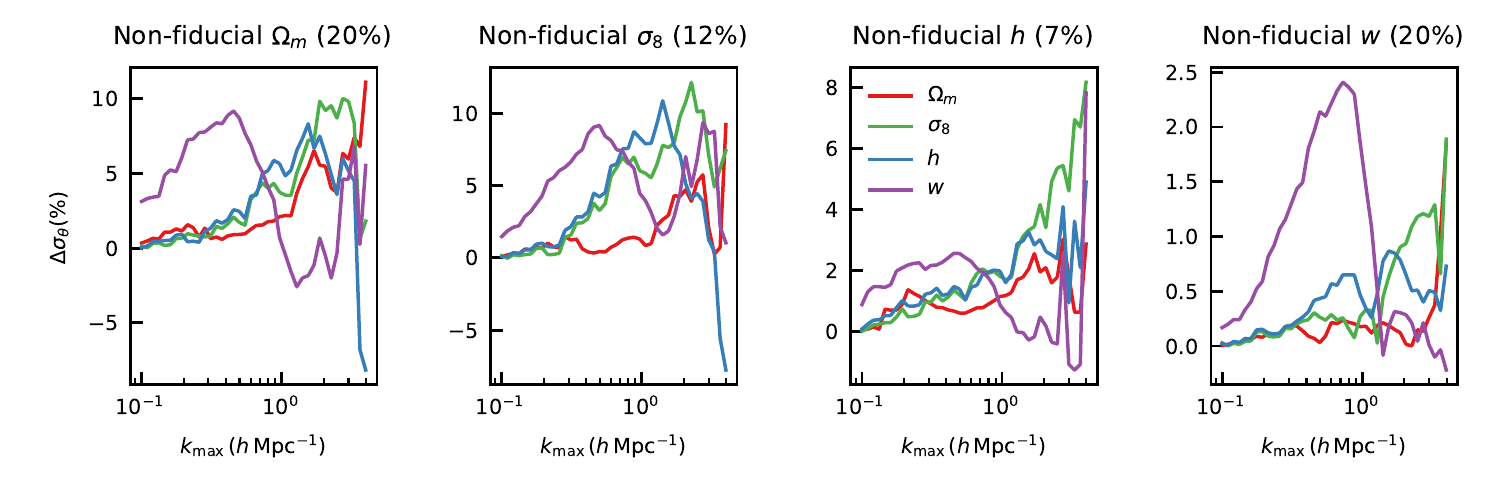}
    \caption{Variation of the 1$\sigma$ error from the Fisher matrix analysis as a function of $\kmax$ assuming the covariance as given by Eq.~(\ref{covtilde}), with the diagonal part set to the fiducial cosmology, while the off-diagonal elements are set to a cosmology with a non-fiducial value of the parameter specified in the title of each panel.}
    \label{fig:param-error-diag}
\end{figure*}

In a more realistic data analysis setting one would sample the likelihood over a large number of points in the cosmological parameter space, while re-computing the whole data covariance at each evaluation. However, this may be unfeasible. To ease this problem one possibility is to model the cosmological dependence of the variance (i.e. the diagonal elements of the covariance) so that it can be varied during the sampling, while keeping the off-diagonal structure of the covariance fixed to a given cosmology. We explore this idea by repeating the Fisher forecast with an approximate covariance given by:
\begin{equation}\label{covtilde}
\tilde{C}_{k_1,k_2}(\boldsymbol{\hat{\theta}})=r_{k_1,k_2}\,\biggr|_{\boldsymbol{\theta}=\boldsymbol{\theta_*}}\; \sqrt{C_{k_1,k_1}C_{k_2,k_2}}\,\biggr|_{\boldsymbol{\theta}=\boldsymbol{\hat{\theta}}},
\end{equation}
where $r_{k_1,k_2}$ is the correlation coefficient:
\begin{equation}
r_{k_1,k_2}=\frac{C_{k_1,k_2}}{\sqrt{C_{k_1,k_1}C_{k_2,k_2}}}. \nonumber
\end{equation}
In this case only the off-diagonal structure of $\tilde{C}$ is computed in a non-fiducial cosmology $\boldsymbol{\theta}_*$, while the diagonal uses the correct parameter values $\boldsymbol{\hat{\theta}}$. This mimics the scenario in which the variance is computed at the cosmology of the sampling points while the correlation coefficient is kept fixed at a given cosmology. The results are shown in Figures~\ref{fig:area-angle-diag}-\ref{fig:param-error-diag}, where we can see that the impact on cosmological parameter errors is now significantly reduced, especially for low $\kmax$ values.

\begin{figure*}
    \includegraphics{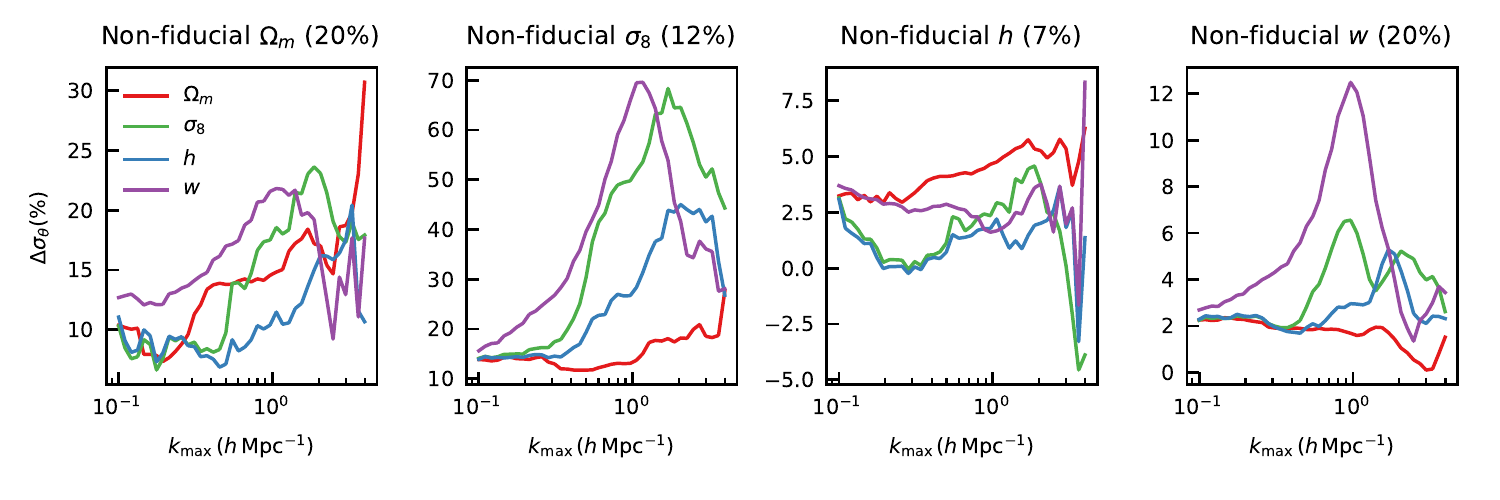}
    \caption{Variation of the 1$\sigma$ error from the Fisher matrix analysis as a function of $\kmax$ assuming the covariance as given by Eq.~(\ref{covparam-error-fix-ng}) with the Gaussian part set to the fiducial cosmology, while the non-Gaussian part is set to a cosmology with a non-fiducial value of the parameter specified in the title of each panel.}
    \label{fig:param-error-fix-ng}
\end{figure*}

In principle, estimating the variance of the matter power spectrum is easier than computing the entire covariance structure, nevertheless it may be still hard to capture all the relevant non-Gaussian contributions to the variance. In such a case, one may think of simplifying the problem by fixing the whole non-Gaussian part of the covariance to a given cosmology and only vary the Gaussian part during the sampling. In this case the approximate covariance is given by:
\begin{multline}\label{covparam-error-fix-ng}
\tilde{C}_{k_1,k_2}(\boldsymbol{\hat{\theta}})=r_{k_1,k_2}\,\biggr|_{\boldsymbol{\theta}=\boldsymbol{\theta_*}}\; \sqrt{C^G_{k_1,k_1}\biggr|_{\boldsymbol{\theta}=\boldsymbol{\hat{\theta}}}+C^{nG}_{k_1,k_1}\biggr|_{\boldsymbol{\theta}=\boldsymbol{\theta^*}}}\\
\sqrt{C^G_{k_2,k_2}\biggr|_{\boldsymbol{\theta}=\boldsymbol{\hat{\theta}}}+C^{nG}_{k_2,k_2}\biggr|_{\boldsymbol{\theta}=\boldsymbol{\theta^*}}},
\end{multline}
where $G$ and $nG$ indicate the Gaussian and non-Gaussian parts respectively. We show the results of the corresponding Fisher analysis in Figure~\ref{fig:param-error-fix-ng}, where we can see that this strategy does not significantly reduce the impact on the parameter errors.

We would like to stress that in a realistic galaxy clustering analysis, the impact of the cosmological dependence of the covariance on the cosmological parameter inference might have a smaller effect than what we have found here. This is because such an analysis will propagate uncertainties on the galaxy bias as well as the effect of shot noise. Whether such nuisance parameters can reduce or absorb the impact of the cosmological dependence of covariance goes beyond the scope of this work and we leave to a future study.

\section{Conclusion}
\label{Sec:Conclusion}
In this work we have investigated the cosmological dependence of the matter power spectrum covariance. To this purpose we have used the DEUS-PUR {\it Cosmo} set of simulations consisting of a large set of independent N-body realizations for different cosmological models characterized by different values of the cosmic matter density $\Omega_m$, the amplitude of matter density fluctuations $\sigma_8$, the reduced Hubble parameter $h$ and the dark energy equation of state $w$. This dataset has enabled us to estimate the covariance matrix for different cosmological parameter values and evaluate its first and second derivatives around a fiducial cosmological model. 
We found that the non-Gaussian part of the covariance from the non-linear clustering of matter exhibits a varying degree of dependence on the different parameters at different redshifts. In particular $\sigma_8$ and $w$ have the largest impact at high redshift, while $\Omega_m$ and $h$ at low redshift. 
The analysis of the covariance derivatives indicates that the convergence of a second order Taylor expansion around the fiducial cosmology to approximate the cosmological dependence of the covariance is rather slow since the first order coefficients of the expansion are of order of unity. In the case of $\sigma_8$, the first order coefficient is larger than unity at high redshift, potentially indicating a non-linear dependence of the covariance on this parameter. The different cosmological model parameters considered here span $\sim 10\%$ variation around the fiducial cosmology, and yet it can lead to important differences in the power spectrum covariance at $\gtrsim100\%$ level on some scales and redshifts.
We have evaluated the impact of the cosmological parameter dependence of the covariance on cosmological parameter inference through a Fisher matrix approach. In particular, we have estimated the parameter uncertainties assuming a non-fiducial model covariance as function of the maximum mode $\kmax$ probed by a galaxy survey. We found significant differences with respect to the case with the covariance set to the fiducial cosmology. The largest effect occurs for non-fiducial values of $\sigma_8$ and $\Omega_m$ with deviations larger than $50\%$ level already at modest $\kmax\sim0.1-0.2 \,\hMpcinv$. On the other hand, the impact on the degeneracy between pair of parameters is less significant, exceeding the $10\%$ level only at $\kmax >1 \,\hMpcinv$. 

These results suggest that the cosmological parameter dependence of the non-Gaussian part of the covariance may impact the cosmological analyses from future surveys of the large scale structures. It is worth emphasizing that the quasi-linear and non-linear scales over which the power spectrum covariance exhibits such a large dependence on the cosmological parameters are also probed by cosmic shear measurements. Hence, it is reasonable to expect that the effects we have found in our analysis may also impact the parameter inference from weak lensing observations. It is important to remind that in our analysis we have neglected the impact of galaxy bias and shot noise. It is yet to be determined whether these nuisance parameters may reduce or enhance the impact of the cosmological dependence of the covariance. However, these depend on the characteristics of the galaxy survey considered, which is beyond the scope of the work presented here.

The current approach of keeping the covariance fixed to the fiducial cosmology when sampling the likelihood is likely to alter the shape of the posterior and consequently introduce systematic uncertainties on the cosmological parameter inference. Since it is not possible to run thousands of simulations to evaluate the covariance for each point in the parameter space that is explored by the likelihood sampling, the cosmological dependence of the covariance need to be modelled. 
Here, we have explored the possibility of modelling such dependence by fixing the off-diagonal part of the covariance matrix to the fiducial cosmology, while letting only the diagonal part vary with cosmology. This significantly reduces the misestimation of the parameter errors from the Fisher analysis, particularly at scales probed by galaxy clustering measurements. The dataset from DEUS-PUR {\it Cosmo} simulations provides an ideal benchmark to test models of the cosmological dependence of the covariance. To this purpose we have made the power spectra used in this work publicly available.

Further investigation is indeed necessary for a more robust assessment of the potential bias induced on the parameter estimation beyond the Fisher matrix approach. Our evaluation of the first- and second-order derivatives of the covariance can provide the foundation for a study that accounts for the non-Gaussian structure of the likelihood, for example using the so called Derivative Approximation for Likelihoods \citep[DALI,][]{2014MNRAS.441.1831S} method. We leave this investigation to a future work. 
 
  \section*{Data availability}
 The power spectra from the DEUS-PUR {\it Cosmo} simulations can be downloaded at \url{https://cosmo.obspm.fr/public-datasets/}.
 
\section*{Acknowledgements}
We would like to thank Jan Kratochvil for his engagement during the initial stages of the project and Romain Teyssier for useful discussion. This work was granted access to HPC resources of IDRIS/CINES through allocations made by GENCI (Grand Equipement National de Calcul Intensif) under the allocations 2016-042287, 2017-A0010402287 2018-A0030402287, 2019-A0050402287. We acknowledge support from the DIM ACAV of the Region Ile de France. The research leading to these results has received funding from the European Research Council under the European Community Seventh Framework Program (FP7/2007-2013 Grant Agreement no. 279954) ERC-StG ``EDECS''. LB acknowledges support from the Starting Grant (ERC-2015-STG 678652) ``GrInflaGal'' of the European Research Council.
\bibliographystyle{mnras}
\interlinepenalty=10000
\bibliography{references} 

\bsp	
\label{lastpage}
\end{document}